\documentclass{article}
\usepackage{spconf,amsmath,graphicx}
\usepackage{cite}
\usepackage{amsmath,amssymb,amsfonts}
\usepackage{algorithmic}
\usepackage{graphicx}
\usepackage{textcomp}
\usepackage{dsfont}
\usepackage{epstopdf}
\usepackage{subcaption}
\usepackage{tikz}
\usepackage{color}
\usepackage{hyperref}
\usepackage{amsbsy}
\usepackage{cleveref}
\usepackage{mathtools}

\def\btheta{{\boldsymbol{\theta}}}

\DeclareMathOperator*{\argmin}{arg\,min}
\DeclareMathOperator*{\argmax}{arg\,max}
\usepackage{algorithmic}

\usepackage[ruled,vlined]{algorithm2e}

\title{Spline Sketches: An Efficient Approach for Photon Counting Lidar}

\name{ Michael P. Sheehan$^1$, Juli\'an Tachella$^2$, Mike E. Davies$^1$\thanks{This work was supported by the ERC Advanced grant, project C-SENSE, (ERC-ADG-2015-694888). Mike E. Davies is also supported by a Royal Society Wolfson Research Merit Award.}}
\address{$^1$ IDCOM, School of Engineering, University of Edinburgh, UK\\
$^2$ CNRS, ENSL, Laboratoire de Physique, Lyon, France}

\begin{document}

\maketitle
\begin{abstract}
Photon counting lidar has become an invaluable tool for 3D depth imaging due to the fine-precision it can achieve over long ranges. However, high frame rate, high resolution lidar devices produce an enormous amount of time-of-flight (ToF) data which can cause a severe data processing bottleneck hindering the deployment of real-time systems. In this paper, an efficient photon counting approach is proposed that exploits the simplicity of piecewise polynomial splines to form a hardware-friendly compressed statistic, or a so-called spline sketch, of the ToF data without sacrificing the quality of the recovered image. As each piecewise polynomial spline is a simple function with limited support over the timing depth window, the spline sketch can be computed efficiently on-chip with minimal computational overhead. We show that a piecewise linear or quadratic spline sketch, requiring minimal on-chip arithmetic computation per photon detection, can reconstruct real-world depth images with negligible loss of resolution whilst achieving $95\%$ compression compared to the full ToF data, as well as offering multi-peak detection performance. These contrast with previously proposed coarse binning histograms that suffer from a highly nonuniform accuracy across depth and can fail catastrophically when associated with bright reflectors. Further, by building range-walk correction into the proposed estimation algorithms, it is demonstrated that the spline sketches can be made robust to photon pile-up effects. The computational complexity of both the reconstruction and range walk correction algorithms scale only with the size of the spline sketch which is independent to both the photon count and temporal resolution of the lidar device.
\end{abstract}

\section{Introduction}\label{SEC: Introduction}
\begin{figure}[h!]
	\centering
	\includegraphics[width=.49\textwidth]{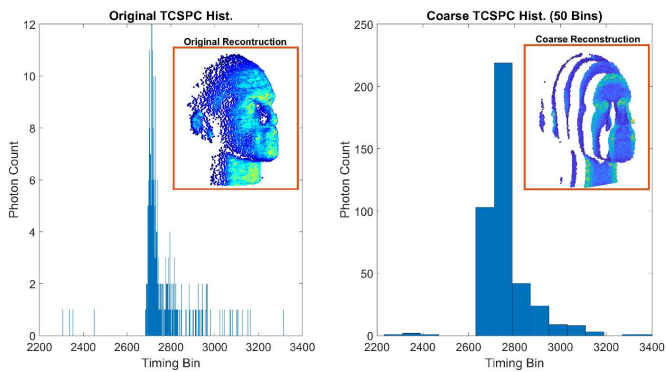}
	\caption{True pixelwise TCSPC histogram and its reconstruction (left) compared with a coarser TCSPC histogram of 50 bins and its reconstruction (right)}
	\label{FIG: Motivation}
\end{figure}

Single-photon light detection and ranging (lidar) is a well-established 3D depth imaging modality that offers millimeter precision\cite{McCarthy:09} over long ranges\cite{Pawlikowska:17}. The technique consists of emitting light pulses and using a single photon avalanche diode (SPAD) to detect the presence of an incoming photon. By using a time-correlated single photon counting system (TCSPC), time-of-flight data can be accumulated over a series of light pulses providing depth and intensity information for the surfaces in each pixel of the scene. A TCSPC histogram is most commonly used to represent the ToF data by clustering the time delay between emitted light pulses and detected photons into time bins discretized over the whole timing depth window period for each pixel in the scene (see \Cref{FIG: Motivation}). Rapid hardware and technological advances in recent years has granted even finer resolution at much quicker frame rates \cite{henderson2018imager,3Dstacked}. Modern lidar devices generate massive amounts of data per second that needs to be transferred off-chip for downstream tasks such as detection, depth estimation and segmentation. As an example, a high rate, high resolution lidar device is capable of imaging a scene containing $512\times512$ pixels at a frame rate of 50 frames per second (fps). Assuming the timing depth window of the laser is discretized over $T=1000$ histogram bins of 16 bit precision, the lidar device would require a data transfer rate of almost 15 GB/s.

\par To tackle the data-transfer bottleneck of modern lidar devices, practitioners have often resorted to using coarse timing bin width in the TCSPC histogram. This technique is often referred to as coarse binning and typically results in a substantial loss of temporal resolution causing a limiting compression-resolution trade-off as demonstrated in~\Cref{FIG: Motivation}. Gyongy et al. \cite{gyongy2020high} proposed a practical solution by increasing the width of the impulse response of the device and using a maximum likelihood estimator to achieve sub-bin resolution with respect to coarse bins. However, as will be demonstrated in \Cref{SEC: Experiments}, we observe that (1) the worse case error is typically highly dependent on the position of the signal with respect to the coarse bins, and (2) if there are highly reflective surfaces (e.g.\ retroreflectors) in the scene, the measured IRF can be much narrower than the systems normal impulse response therefore sub-bin accuracy with respect to the coarse bins cannot be achieved. Recently, Sheehan et al. \cite{sheehan2021sketchedlidar,sheehan2021sketcheddetection,tachella2022SRT3D} proposed a novel solution to tackle the data-transfer bottleneck whilst circumventing the compression-resolution trade-off. The technique consists of forming a compact representation, a so-called sketch, of the ToF data that retains sufficient salient information to accurately estimate the depth and intensity of multiple surfaces in the scene. Fundamentally, it was demonstrated that the size of the sketch only needs to be of the order of the number of surfaces in the scene to achieve negligible loss of resolution\cite{sheehan2021sketchedlidar}. As there are typically no more than 1 or 2 surfaces present per pixel, a substantial compression rate can be achieved. Furthermore, the reconstruction algorithms are based solely on the sketch and admit both a computational and memory complexity that scales only on the size of the sketch and is independent of both the resolution of the lidar device as well as the number of photons detected. However, the method proposed in\cite{sheehan2021sketchedlidar} requires computing several sinusoidal functions for each photon detection which may be challenging to execute within the limited dead-time (a period of insensitivity after a photon detection during which the SPAD cannot register any incoming photons) of the lidar device.
\par In this paper, we introduce new sketch statistics that both captures the necessary salient information in the data while requiring low on-chip arithmetic complexity. To this end we propose a spline sketch approach for photon counting that unites the substantial data transfer compression capabilities of the Fourier sketched lidar technique with the hardware friendly computation of a traditional TCSPC histogram. By replacing the sinusoidal functions of the original sketch in \cite{sheehan2021sketchedlidar} by elementary piecewise polynomial spline functions, we can construct a compressed representation of the ToF data that is almost as statistically efficient as the originally proposed Fourier sketch while having minimal computational overhead on-chip. The main contributions of this paper are:
\begin{itemize}
    \item Through the introduction of piecewise polynomial splines, we design low-cost, hardware-efficient sketches that retain sufficient salient information required to estimate the depth and intensity parameters of the lidar observation model and bridge the gap between coarse binning representations and Fourier sketches whose performance is position independent by definition.
    \item Both closed-form (for linear splines) and iterative algorithms are proposed that can accurately estimate the depth and intensity parameters of the scene while exhibiting a computational and memory complexity that scales solely with the size of the spline sketch.
    \item We show that the performance of coarse binning is highly dependent on the position of the reflector within a given bin. In contrast, linear and quadratic polynomial spline sketches exhibit only mild variation in such performance and achieve theoretical error almost equivalent to the Fourier sketch.
    \item Finally, we show that the spline sketches carry sufficient information to enable the correction of range walk associated with bright reflective surfaces and present a prototype spline sketch range walk correction algorithm.
\end{itemize}
\par The paper is organized as follows: In \Cref{SEC: Background} the concepts of sketched lidar and piecewise polynomial splines are introduced. In \Cref{SEC: Spline Sketch}, the cost of computing the spline sketches on-chip is discussed while the spline sketch based reconstruction algorithms are proposed in \Cref{SEC: Spline Sketch Reconstruction}. Both theoretical Cram\'er-Rao error bounds and the relative performance of the proposed algorithms are demonstrated in Section \ref{SUBSEC: CRB} and \ref{SEC: Experiments}, respectively. Finally, we conclude the paper in \Cref{SEC: Conclusion}.

\section{Background}\label{SEC: Background}
\subsection{Lidar Observation Model}\label{SUBSEC: Oberservation model}
During the course of $N$ laser pulses, the SPAD registers $n\leq N$ photon detections for a given pixel in the scene, where each photon has a detection time denoted by $x_j$ for $1\leq j \leq n$. Assuming there are $K$ distinct surfaces in the field of view each at a timing depth of $t_k\in[0,T-1]$, then a simple model for the detection time $x_j$ of a single photon for a given pixel is given by a mixture distribution \cite{altmann2}\footnote{This model does not capture all aspects of the SPAD device. However it serves as a useful idealisation. The sketching framework is not reliant on this specific model and indeed, a more complicated one will be considered in~\Cref{SUBSEC: Range Walk Correction} when discussing range walk effects.}
\begin{equation}
    \label{Eqn: Observ model}
    \pi(x_j|\btheta)= \sum^K_{k=1}\alpha_k\pi_s(x_j|t_k)+\alpha_0\pi_b(x_j),
\end{equation} 
where $\alpha_1,\alpha_2,\dots,\alpha_K$ denote the probability of detection from the $k$th surface, $\alpha_0$ denotes the probability of detection from an ambient background source (e.g.\ sunlight) and $\sum_{k=0}^K\alpha_k=1$. Given that the lidar system has a known impulse response function (IRF) $h(t)$ 
then the distribution of photons originating from a surface is defined as $\pi_s(x|t)=h(x-t)/H$ where $H=\sum_{t=0}^{T-1} h(t)$. The distribution of background photon detections can be modelled uniformally over the whole timing depth window $\pi_b(x)=1/T$. As a result, single photon counting lidar reduces to estimating the parameters $\btheta=(\alpha_0,\alpha_1,\dots,\alpha_K,t_1,\dots,t_K)$ given a collection of ToF data (e.g.\ TCSPC histogram) for each pixel in the scene. As discussed in \Cref{SEC: Introduction}, the complexities of transferring ToF data are often dependent on the depth resolution of the lidar device which is determined by the discrete number of time bins $T$ used over the whole timing depth window. The memory requirement for a single pixel in a frame is typically $\mathcal{O}(Tb)$ where $b$ denotes the integer bit precision which can be memory expensive in modern day devices with fine depth resolution.
\subsection{Sketched Lidar}\label{SUBSEC: Sketched Lidar}
Recently, Sheehan et al. \cite{sheehan2021sketchedlidar,tachella2022SRT3D} developed a novel solution to capture sufficient salient information from the ToF data $\{x_j\}^n_{j=1}$ to estimate the parameters of the observation model in (\ref{Eqn: Observ model}). The TCSPC histogram (see \Cref{FIG: Motivation}) can be seen as a statistic that captures enough information from the ToF data required for estimation. However it is a very inefficient representation with memory size scaling proportional to required resolution.
The authors in \cite{sheehan2021sketchedlidar} designed a much more compact representation of the ToF data that has a size independent of the resolution $T$ of the lidar device and the number of photon detections $n$.
\par The basic concept of sketching is to introduce a nonlinear mapping $\Phi(x)$ that defines a sketch statistic as its empirical average, $\mathbf{z} = \frac{1}{n}\sum_{j=1}^n \Phi(x_j)$ with respect to data $x_j$ drawn from a  probability density $\pi(x|\theta)$. Then we can estimate the parameters $\theta$ by comparing the empirical sketch, $\mathbf{z}$ with the hypothesised expectation: $\mathbb{E}_{x\sim \pi(\cdot|\theta)}\Phi(x)$. In the original sketched lidar work\cite{sheehan2021sketchedlidar} a particular Fourier sketch proposed that was inspired by compressive learning theory \cite{CompressiveLearning}. Let $\Phi_\mathcal{F}(x)=[e^{ {\rm i}\omega_{\ell}x}]_{l=1}^{m}$ denote $m$ complex Fourier features, then the Fourier sketch of size $m$ can be defined as 
\begin{equation}
\label{Eqn: Fourier Sketch}
    \mathbf{z}_\mathcal{F} = \frac{1}{n}\sum_{j=1}^n \Phi_\mathcal{F}(x_j).
\end{equation}
Due to the summation in (\ref{Eqn: Fourier Sketch}), the sketch can be updated throughout the pulse cycle for each photon detection. Once the sketch has been computed, the parameters of the observation model in (\ref{Eqn: Observ model}) can be estimated, e.g.\ through a generalized method of moments \cite{gemmhall} scheme
\begin{equation}
\label{Eqn: GeMM Fourier Sketch}
  \argmin_{\btheta} \lVert \mathbf{z}_\mathcal{F} - \mathbb{E}_{x\sim \pi}\Phi_\mathcal{F}(x) \rVert_\mathbf{W}^2,
\end{equation}
for some positive definite weighting matrix $\mathbf{W}\in\mathbb{C}^{m\times m}$. Advantageously for the Fourier sketch, the sketch in expectation is simply the characteristic function (CF) of the observation model in (\ref{Eqn: Observ model}) sampled at the $m$ frequencies $\omega_1,\dots,\omega_m$. The CF exists for all distributions and, in the case of the observation model in (\ref{Eqn: Observ model}), has a simple closed form solution which can be easily computed. Fundamentally, the size of the sketch (i.e.\ the number of CF samples) required to accurately estimate $\btheta$ needs only to be of the order of the number of surfaces in the pixel, for instance $m=\mathcal{O}(K)$. Moreover, in\cite{sheehan2021sketchedlidar,tachella2022SRT3D}, the authors propose several sketch-only reconstruction algorithms, including exploiting powerful spatial denoisers, that have both a memory and computational complexity of $\mathcal{O}(m)$.
\par One of the challenges with the original sketched lidar approach is that the feature function $\Phi_\mathcal{F}$ requires nontrivial on-chip processing. Firstly, the feature function in (\ref{Eqn: Fourier Sketch}) consists of $2m$ sinusoidal functions\footnote{The original Fourier sketch is a complex number, however it is possible to exploit the complex structure to simultaneously calculate sin and cos within FPGA implementations.} which is resource intensive, despite the existence of various FPGA implementations, e.g.\ \cite{deDinechin2014}. Secondly, the Fourier sketch is a \textit{global} sketch as each sinusoidal function has a support equal to the whole pulse cycle and therefore all $2m$ components of the sketch must be updated for each photon detection. This is in contrast to the TCSPC histogram where only a single timing bin needs to be updated for each photon detection. In this paper we propose to replace the Fourier features $\Phi_\mathcal{F}$ with a low-cost, semi-local spline sketch consisting of $M$ piecewise polynomial spline feature functions that have a limited support over the timing depth window. 
While one can intuitively think of the spline functions as approximating the Fourier representation, it is important to stress that they define valid sketches in their own right (we are not simply approximating the Fourier sketch) and as we will see in~\Cref{SUBSEC: Closed Form Sol} have additional pleasing properties that are unrelated to the previous Fourier sketch. 
\subsection{Cardinal Basis Splines}\label{SUBSEC: Piecewise Poly Spline}
\begin{figure}[h!]
	\centering
	\includegraphics[width=.49\textwidth]{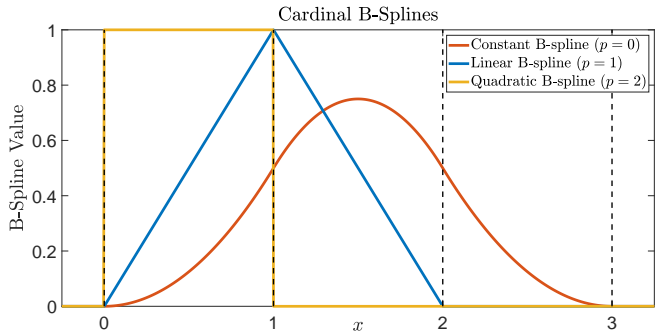}
	\caption{The $p=0,1$ and 2 degree cardinal B splines.}
	\label{FIG: Cardinal B splines}
\end{figure}

Cardinal splines are piecewise polynomial functions of degree $p$ that are used extensively in approximation theory \cite{unser1}. Let $\phi_0$ denote the spline of degree $p=0$ defined as 
\begin{equation}
\label{Eqn: Cardinal 0 B spline}
\phi_0(x)=\begin{cases} 
      1 & x\in[0,1) \\
      0 & \text{otherwise}.
\end{cases}
\end{equation}
Then the degree $p$ spline $\phi_p$ is defined as the $p+1$ convolution of $\phi_0$:
\begin{equation}
\label{Eqn: B spline convolution}
    \phi_p(x) = (\underbrace{\phi_0*\phi_0*\dots*\phi_0}_{p+1 \text{ times}})(x)
\end{equation}
where $(f*g)(t)=\int_\mathbb{R}f(t-s)g(s)\,ds$ denotes the convolution operator. In this paper, we will only consider the first three degree splines due to their overall computational simplicity. It can be easily deduced from (\ref{Eqn: Cardinal 0 B spline}) that the linear ($p=1$) and quadratic ($p=2$) cardinal splines are given as 

\begin{equation}
\label{Eqn: Cardinal 1 B spline}
\phi_1(x)=\begin{cases} 
      x & x\in[0,1) \\
      2-x & x\in[1,2) \\
      0 & \text{otherwise}
\end{cases}
\end{equation}
\noindent and

\begin{equation}
\label{Eqn: Cardinal 2 B spline}
\phi_2(x)=\begin{cases} 
      \frac{1}{2}x^2 & x\in[0,1) \\
      \frac{1}{2}\;+\;(x-1)\;-\;(x-1)^2 & x\in[1,2) \\
      \frac{1}{2}\;-\;(x-2)\;+\;\frac{1}{2}(x-2)^2 & x\in[2,3) \\
      0 & \text{otherwise}
\end{cases}
\end{equation}
respectively. Cardinal splines of degree $p$ have the appealing properties that (1) they are $p$ times differentiable and (2) are compactly supported in $\phi_p\in[0,p+1]$. \Cref{FIG: Cardinal B splines} depicts the first three degree cardinal splines.

\section{Spline Sketch}\label{SEC: Spline Sketch}
\begin{figure}[h!]
	\centering
	\includegraphics[width=.49\textwidth]{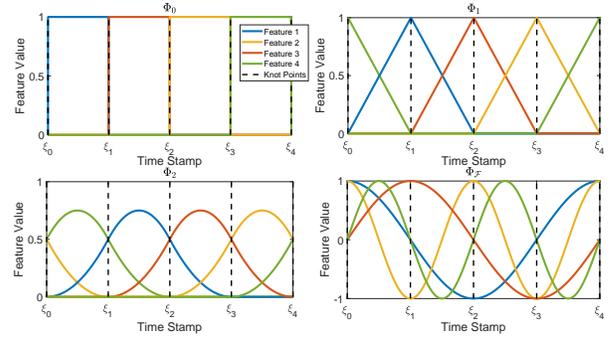}
	\caption{The $p=0,1,2$ spline and Fourier sketches of size $M=4$. The dotted black lines illustrate the partition of the timing depth window into 4 knot intervals defined by the knot points $\xi_i$. }
	\label{FIG: Compare Features}
\end{figure}

In this paper, we are interested in constructing a sketch consisting of several splines that together cover the whole support of the timing depth window $T$, which we will treat as having periodic boundary conditions, exactly as in the Fourier setting. To do so, let $\xi_0,\xi_1,\dots,\xi_M$ be a sequence of equispaced points, called knot points, over the whole timing depth window $T$ such that $\xi_0=0$ and $\xi_M=T$. In addition, let $\Delta\coloneqq T/M$ denote the knot interval distance between two consecutive knot points. Then the scaled and translated spline of degree $p$ is defined as
\begin{equation}
    \label{Eqn: Scaled and Translated B Spline}
    \phi_{i,p}(x) \; = \; \phi_p\left(\frac{x}{\Delta}-i\right),
\end{equation}
for $i=0,1,\dots,M-1$. The spline $\phi_{i,p}$ is supported on $\left[i\Delta,(i+p+1)\Delta\right]$. As a result, a spline sketch of size $M$ can be expressed as 
\begin{equation}
    \label{Eqn: B spline sketch}
    \mathbf{z}_p = \frac{1}{n}\sum^n_{i=1}\Phi_p(x_i)
\end{equation}
where $\Phi_p(x) = \left[\phi_{i,p}(x)\right]_{i=0}^{M-1}$ is the set of spline feature functions of degree $p$. \Cref{FIG: Compare Features} illustrates the spline feature function $\Phi_p$ of size $M=4$ for $p=0,1,2$ as well as the real and imaginary components of the original Fourier sketch. 
\par One can see straightaway that for $p=0$ the spline feature function is equivalent to forming $M$ coarse bins over the whole timing depth window. As the degree $p$ increases, we observe two important points (1) the support of each spline feature increases linearly with respect to $p$, (2) each knot interval $\Delta$ contains $p+1$ spline features. These two observations will become important when we consider the on-chip implementations of the spline sketches in \Cref{SUBSEC: Logic Implementation}. Each of the Fourier features ($\phi_{\omega_j}(x)=e^{{\rm i}\omega_j x}$) are depicted in the bottom right of \Cref{FIG: Compare Features}. In contrast to the spline features, each individual Fourier feature has a support equal to the whole timing depth window $T$.
\subsection{Hardware Implementation}\label{SUBSEC: Logic Implementation}
Algorithm \ref{Alg: Online proccessing} details a prototype on-chip algorithm for computing a spline sketch of degree $p$ in an online manner. Once the appropriate knot interval of the photon detection has been established, only $p+1$ of the total $M$ sketch features need to be updated. We detail further the total number of arithmetic operations required per photon detection for each of the $p=0,1,2$ spline sketches. 

\begin{algorithm}
\caption{Sketch Online Processing}
\label{Alg: Online proccessing}
\begin{algorithmic}
\STATE \textbf{Initialisation:} Degree $p$ , $\mathbf{z}_p=0, n=0$
\WHILE{Acquisition Window}
\IF{New Photon Arrival with Time Stamp $x$}
\STATE Detect Knot interval $x\in\left[i\Delta,(i+1)\Delta\right)$
\FOR{$q \xleftarrow{} 0 \text{ to } p$}
\STATE $\mathbf{z}_p = \mathbf{z}_p + \phi_{i-q,\Delta,p}(x)$
\ENDFOR
\STATE $n\xleftarrow{}n+1$
\ENDIF
\ENDWHILE
\STATE $\mathbf{z}_p \xleftarrow{}\mathbf{z}_p/n$
\ENSURE The sketch $\mathbf{z}_p$ is transferred off-chip for post-processing.
\end{algorithmic}
\end{algorithm}

For efficiency it makes sense to set $T$ and $M$ to integer powers of 2 so that identifying the knot interval can be read from the $\log(M)$ leading bits of the $\log(T)$-bit digital representation of the detection time. The remaining $b$-bits then represent the relative position within the knot interval which we denote by $r$. We further note that identifying the knot interval is the same for $p = 0, 1$ and $2$. Therefore in examining the different arithmetic operations we will only focus on the calculation of accumulated spline function values. For the spline function calculation we assume that the spline functions are suitably scaled such that exact integer arithmetic can be used throughout. For the $p=0$ degree spline function $\Phi_0$ (i.e.\ coarse binning), once we have located the appropriate bin all that is required is to update the bin count which involves a single one bit addition (e.g.\ implemented via a ripple counter). In the case of the $p=1$, two neighboring sketch values have to be calculated. From (\ref{Eqn: Cardinal 1 B spline}) we can see the first value is simply $r$ and requires no further calculation. The second value is $2^{b+1} - r$ and requires a single addition. These values then have to be added to the current sketch values requiring a further two additions. Therefore the linear spline needs to execute a total of 3 addition/subtractions per photon detection. 
\par For $p=2$, there are three spline values that must be calculated per photon detection. In terms of the relative knot interval position, $r$, the three values that require calculation are: $r^2$,    $2^{2b}-2^b r-r^2$, and $2^{2b} - 2^{b+1} r + r^2$. Notice here that all scaling involved is by powers of 2 and therefore only involves a binary shift (whose computation we ignore). Once calculated, these values will need to be added to the associated three sketch components. Therefore the quadratic spline needs to execute a total of 1 multiplication and 7 addition/subtractions per photon detection. In contrast, while there are various ways to compute the original Fourier sketch in (\ref{Eqn: Fourier Sketch}), e.g.\ CORDIC or local polynomial approximation\footnote{i.e.\ $\sin(\omega_j x)\approx \omega_j x-\frac{\omega_j^3x^3}{6}$ and $\cos(\omega_j x)\approx 1 - \frac{\omega_j^2x^2}{2}+\frac{\omega_j^4x^4}{24}$ for $j=1,2,\dots, M/2$.}\cite{deDinechin2014}, depending on the latency and/or the computation desired, the computational cost is typically much higher. Furthermore, in each case all $M$ sketch values need to be updated per photon detection.  Specific details of the arithmetic operations required for each spline sketch per photon detection is summarized in \Cref{TAB: operations of sketches}.

\begin{table}[h!]
\caption{Table highlighting the arithmetic operations required for each sketch per photon detection.}
\label{TAB: operations of sketches}
\begin{tabular}{|c|c|c|c|c|}
\hline
Feature Function & Active Features & Add/Sub & Mult.           \\ \hline
$\Phi_0$     & 1               & 1 (1-bit)       & 0        \\ \hline
$\Phi_1$       & 2               & 3       & 0              \\ \hline
$\Phi_2$      & 3               & 7       & 1               \\ \hline
\end{tabular}
\end{table}

\section{Spline Sketch Reconstruction}\label{SEC: Spline Sketch Reconstruction}
\subsection{Closed Form Solution}\label{SUBSEC: Closed Form Sol}
\begin{figure}[h!]
	\centering
	\includegraphics[width=.49\textwidth]{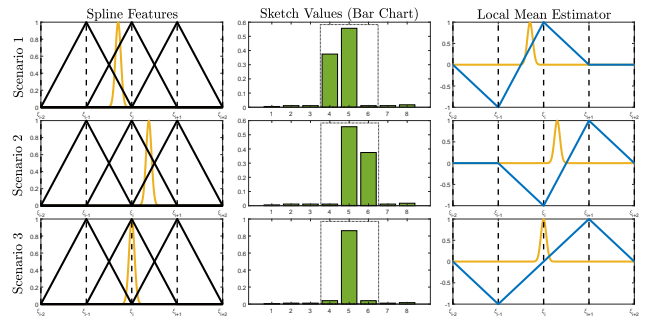}
	\caption{Left column: the 3 (out of 8) spline sketches in the region of the signal. Middle column: a bar chart of the spline sketches. Left column: local means estimator of each scenario. }
	\label{FIG: closed form solution}
\end{figure}

We begin by proposing a simple closed form solution for the piecewise linear spline sketch\footnote{In principle this could also be done for p=2 but we have not pursued this here} that accurately estimates the parameters of the observation model in (\ref{Eqn: Observ model}). The method assumes an IRF with local support and utilises adjacent linear splines located in the region of the reflected pulse to form a local mean estimator. Due to the semi-local nature that is inherent to splines (see \Cref{SEC: Spline Sketch}) one can easily establish a subset of splines that are needed to construct the local mean estimator. First we locate the individual spline sketch component with the maximum magnitude, 
\begin{equation}
    l\; =\; \argmax_{1\leq j\leq M}\,\mathbf{z}^j_1
\end{equation}
where $\mathbf{z}^j_1$ is the $j$th entry of the sketch $\mathbf{z}_1$. Given that the largest individual sketch $\mathbf{z}_1^l$ has been located and assuming that the support of the IRF is less than one knot interval, then there are only 3 possible scenarios that can exist. \Cref{FIG: closed form solution} depicts each of the 3 possible scenarios.

\par In scenario 1, the IRF is contained within 1 knot interval on the left side of the $l$th spline. This is reflected in the bar chart (middle) of the spline sketch values. Similarly, in scenario 2, the IRF is contained within 1 knot interval on the right side of the $l$th spline. This is also reflected in the bar chart of the spline sketch values where the second largest sketch value is to the right of $\mathbf{z}_1^l$. Finally in scenario 3, the IRF straddles a knot point and is contained within both knot intervals, as captured by the bar chart of the spline sketch values. In each scenario, we can build a simple local mean estimator by taking the difference of adjacent sketch values about $\mathbf{z}_1^l$ as is captured in the third column in \Cref{FIG: closed form solution}. Notice that when the IRF straddles a knot point, the local mean estimator is calculated by the difference between $\mathbf{z}_1^{l-1}$ and $\mathbf{z}_1^{l+1}$ hence the estimator is linear over 2 knot intervals. Specifically, the 3 estimators are calculated as follows
\begin{align}
\label{Eqn: closed form depth estimates}
    \text{Scenario 1:} &\quad \hat{t} \; =\; \xi_{l-1}+\frac{\Delta}{2}\;+\; \frac{T(\mathbf{z}^l_1-\mathbf{z}^{l-1}_1)}{2M\alpha}\\
     \text{Scenario 2:} & \quad \hat{t} \; =\; \xi_{l}+\frac{\Delta}{2}\;+\; \frac{T(\mathbf{z}^{l+1}_1-\mathbf{z}^{l}_1)}{2M\alpha}\\
       \text{Scenario 3:} & \quad \hat{t} \; =\; \xi_{l}\;+\; \frac{T(\mathbf{z}^{l+1}_1-\mathbf{z}^{l-1}_1)}{M\alpha}
\end{align}
In each case, the depth estimator $\hat{t}$ is dependent on knowing or accurately estimating $\alpha$. However,this can easily be estimated from neighbouring spline values that are assumed to only contain background photons. For the features located in positions that consist solely of photon detections from ambient sources it can be easily seen (see (\ref{Eqn: Observ model})) that 
\begin{equation}
    \mathbb{E}_{x\sim\pi} \, \phi_{i,p}(x) \; = \; \frac{1-\alpha}{M}.
\end{equation}
One can therefore use the sketch values $\mathbf{z}^i_1$ that are known to only contain background counts to estimate $\alpha$ as follows:

\begin{equation}
\label{EQN: first noise corrected eqn}
    \hat{\alpha} \; = \; 1- M\mathbf{z}^i_1.
\end{equation}
If one has access to multiple sketch values that contain photon detections solely from ambient sources then taking the average over those estimates can lead to a more accurate $\hat{\alpha}$.
\par Given the 3 estimators above, one can choose the scenario which produces the lowest spline sketch loss:
\begin{equation}
\label{Eqn: Spline 1 Sketch loss} 
    \lVert \mathbf{z}_1 -\mathbb{E}_{x\sim\pi} \Phi_1(x)\rVert_2.
\end{equation}
\noindent As one has to only compute the loss in (\ref{Eqn: Spline 1 Sketch loss}) a total of 3 times, the computational complexity of this closed form solution is minimal. As the method returns the (local) mean of the IRF distribution, it is independent of the IRF shape (incorporating the appropriate leading-edge to mean correction as appropriate). For additional accuracy one could further fine-tune the estimates by running a few iterations of a spline version of the SMLE algorithm proposed in \cite{sheehan2021sketchedlidar} or additionally incorporate powerful spatial denoisers as proposed in \cite{tachella2022SRT3D}. We evaluate the performance of this method in \Cref{SEC: Experiments} and compare the error it produces to the optimal Cram\'er-Rao lower bound.

\subsection{Matching Pursuit Algorithm}
The closed form estimator proposed in \Cref{SUBSEC: Closed Form Sol} produces a simple and computationally cheap solution. However, 
the solution is dependent on an accurate estimate of the reflectivity $\alpha$ which can become problematic if there are multiple reflectors in a scene. In the case of multiple reflectors per pixel, we propose a robust matching pursuit (MP) inspired algorithm, similar to the one proposed by Keriven et al. in \cite{keriven2018sketching}, that iteratively selects a reflector that is most correlated to the current residual error. In addition, it can comfortably handle spline sketches of any degree $p$. Details of the proposed algorithm are given in Algorithm 2.

\begin{algorithm}
\caption{Spline sketch matching pursuit reconstruction algorithm.}
\label{Alg: MP Recon alg}
\begin{algorithmic}
\STATE \textbf{Initialisation:} degree $p$, number of reflectors $K$, $\theta = \emptyset$, residual $\mathbf{r}=\mathbf{z}_p$ 
\FOR{$k \xleftarrow{} 1 \text{ to } K$}
\STATE Find a reflector highly correlated with the residual
$t^*_k \xleftarrow{}\,\max_t \Big\langle\frac{\mathbb{E}_{x\sim \pi_s(t)}\,\Phi_p(x)}{\lVert\mathbb{E}_{x\sim \pi_s(t)}\,\Phi_p(x)\rVert_2},\mathbf{r}\Big\rangle$\newline
\STATE  $\mathbf{z}_{\theta_k}\,\xleftarrow{}\,\mathbb{E}_{x\sim \pi_s(t^*_k)}\,\Phi_p(x)$\newline
\STATE $\alpha_k\,\xleftarrow{}\,\frac{\langle\mathbf{z}_p,\mathbf{z}_{\theta_k}\rangle}{\lVert \mathbf{z}_{\theta_k}\rVert_2}$\newline
\STATE  $\theta \xleftarrow{} \theta \cup \left\{\alpha_k,t_k\right\}$\newline
\STATE $\mathbf{r}\xleftarrow{}\mathbf{z}_p - \alpha_k\mathbf{z}_{\theta_k}$
\ENDFOR
\STATE  Normalise $\alpha$ such that $\sum_{k=0}^K\alpha_k=1$.
\ENSURE Estimates for the depth and intensity information of $K$ reflectors in a given pixel.
\end{algorithmic}
\end{algorithm}

\par The computational complexity of the MP spline sketch reconstruction algorithm is $\mathcal{O}(TMK)$ and only takes $K$ iterations to complete. Due to the semi-local nature of the spline sketches, one can easily initialise $t_k$ with a coarse grid in the region of the maximum spline $\mathbf{z}_p^{l}$ (see the middle column of \Cref{FIG: closed form solution}) and in the specific case of linear splines, one can use the simple closed form solution discussed in \Cref{SUBSEC: Closed Form Sol} for a good initialisation. In doing so, the computational complexity reduces to $\mathcal{O}(T'MK)$ where $T'\leq T$ is the size of the local coarse grid. The MP algorithm can handle empirical IRF functions by approximating $\mathbb{E}_{x\sim \pi}\,\Phi_p(x)$ using simple numerical integration schemes, for instance the trapezium rule \cite{NumericalIntegration}, which can be stored within a look-up table. One can further fine-tune the estimates by running a few iterations of a spline version of the SMLE algorithm proposed in \cite{sheehan2021sketchedlidar} or additionally incorporate powerful spatial denoisers as proposed in \cite{tachella2022SRT3D}. In \Cref{SEC: Experiments}, we evaluate the MP algorithm and compare the error it produces to the optimal Cram\'er-Rao lower bounds.

\subsection{Range Walk Correction}\label{SUBSEC: Range Walk Correction}
Lidar systems are subject to systematic range errors that arise due to a difference between the estimated results and the ground truth distance of the object in the scene \cite{Tontini20,heide2018sub}. This can happen due to many aspects of the detector's response, but it is particularly linked to imaging highly reflective objects such as a retroreflector. Retroreflectors are ultra reflective surfaces that lead to a high probability of photon detection per laser pulse. Due to the intrinsic deadtimes of the lidar device, the SPAD detection deviates from the classical Poisson statistical model and more sophisticated models need to be considered. Here we will use a model similar to the one presented in\cite{heide2018sub} to examine the associated range walk effects and possible sketch corrections. The photon flux $\lambda_i$ on the SPAD during the $i$th time bin of the timing depth window, denoted by the interval $I_i$, can be modelled by an inhomogeneous Poisson process with rate function
\begin{equation}
    \lambda_i \; = \; \int_{I_i} \mu \left(\beta\frac{h(x-t)}{H} + s\right)\; dx
\end{equation}
\noindent where $\mu$ is the SPAD photon detection probability ($\sim 1\%$), $\beta \in [0,1]$ is the target reflectivity and $s$ is the level of detections from background sources. In the medium and large flux regimes, the SPAD tends to detect photon arrivals at earlier bins along the timing depth window, resulting in an observed IRF that is distorted towards the leading edge of the original pulse. This phenomena is commonly referred to as \textit{photon pile-up} and can cause large range errors for highly reflective surfaces. 

\begin{figure}[h!]
	\centering
	\includegraphics[width=.49\textwidth]{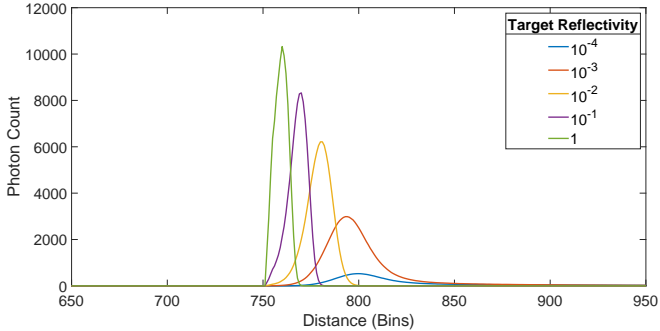}
	\caption{Distorted IRFs as a function of the target reflectivity.}
	\label{FIG: distorted IRFs}
\end{figure}

\par
\begin{figure}[h!]
	\centering
	\includegraphics[width=.49\textwidth]{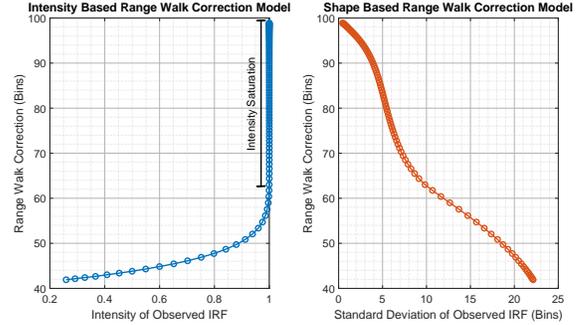}
	\caption{Left: A look-up table evaluating intensity to range walk error. Right: A look-up table evaluating the standard deviation of the observed IRF to range walk error.}
	\label{FIG: RW LUT tables for both}
\end{figure}

 \Cref{FIG: distorted IRFs} depicts the distortion of a real life IRF from the face dataset in \cite{altmann2016lidar} simulated using the multinomial distribution in \cite{heide2018sub} for different target reflectivity $\beta$ and with a SPAD photon detection probability of $\mu=0.01$. It can be seen that as the target reflectivity increases to 1, photon detections accumulate more frequently towards the rising edge of the original IRF with larger intensity. As the location of the reflecting surface can be identified with the leading edge of the observed IRF, any estimator based on the peak or the mean of the distribution will need a reflectivity dependent correction, often called range walk correction, in order to accurately locate the leading edge.  Intensity based correction models are a well-established technique in range walk correction \cite{Tontini20,RW1} and can be easily integrated into a spline sketch estimation algorithm as any degree spline sketch contains intensity information. The left of \Cref{FIG: RW LUT tables for both} shows the range walk correction as a function of the normalized intensity calculated using a spline sketch of size $M=50$. As with all intensity based range walk models, the range walk error can be uniquely corrected up until a point of saturation. This saturation occurs when the probability of detection is essentially 1 (i.e.\  $n\approx N$) and happens between the medium and high flux regimes. After the saturation point has been hit, it is impossible to distinguish between different range-walk errors using only intensity which can lead to large estimation errors (as it will be demonstrated in \Cref{FIG: RW RMSE}).

\par In order to accurately correct for range walk within the saturated regime, it is necessary to capture higher order statistics related to the shape of the observed IRF. This can be easily done using a quadratic spline sketch as they also capture local second order statistics of the ToF data allowing one to distinguish between the variances of the IRFs. It can be seen in \Cref{FIG: distorted IRFs} that as the target reflectivity increases, the observed IRF becomes ever more concentrated about the leading edge. The right of \Cref{FIG: RW LUT tables for both} shows the range walk error as a function of the local estimate of the standard deviation which can be calculated using the quadratic spline sketch ($M=50$) following the same principle as the local mean estimators presented in \Cref{SUBSEC: Closed Form Sol}. In contrast to the intensity based range walk model, the shape based range walk model defines a robust look-up table correction over the full range of reflectivities. We therefore have a unique and well-defined correction scheme. 
\par In a similar fashion to the closed form estimator in \Cref{SUBSEC: Closed Form Sol}, one can construct a local standard deviation estimator in closed form that is both simple and computationally cheap. First we locate the individual spline sketch with the maximum magnitude

\begin{equation}
    l\; =\; \argmax_{1\leq j\leq M}\,\mathbf{z}^j_2
\end{equation}
where $\mathbf{z}^j_2$ is the $j$th entry of the quadratic spline sketch $\mathbf{z}_2$. Next, we remove the noise contribution from the individual sketches by defining the noise corrected sketches as 
\begin{equation}
    \label{EQN: RW noise corrected sketches}
    \Tilde{\mathbf{z}}^j_2 \; = \; \frac{1}{\alpha}\left(\mathbf{z}^j_2 - \frac{1-\alpha}{M}\right).
\end{equation}
By letting 
\begin{equation}
    \Tilde{\mathbf{z}}_\text{sub} \; = \; \left[ \Tilde{\mathbf{z}}^{l-2}_2, \Tilde{\mathbf{z}}^{l-1}_2,\Tilde{\mathbf{z}}^{l}_2,\Tilde{\mathbf{z}}^{l+1}_2,\Tilde{\mathbf{z}}^{l+2}_2\right]
\end{equation}
define the collection of noise corrected sketches in the neighbourhood of $\mathbf{z}_2^l$, then it can be easily shown that the depth and shape estimates have the closed form solution
\begin{equation}
    \hat{t} \; = \; \xi_l + \mathbf{c}_1^T\Tilde{\mathbf{z}}_\text{sub},
\end{equation}
and
\begin{equation}
    \hat{\sigma}^2 \; = \;  \mathbf{c}_2^T\Tilde{\mathbf{z}}_\text{sub} \;-\; \left(\mathbf{c}_1^T\Tilde{\mathbf{z}}_\text{sub}\right)^2,
\end{equation}
\noindent respectively, where
\begin{align*}
    \mathbf{c}_1 & = [-2\Delta, -\Delta, 0,\Delta,2\Delta],\\
    \mathbf{c}_2 & = \left[\frac{15}{4}\Delta^2,\frac{3}{4}\Delta^2, -\frac{1}{4}\Delta^2,\frac{3}{4}\Delta^2,\frac{15}{4}\Delta^2\right]
\end{align*}
and recalling that $\Delta = \frac{T}{M}$.

Here we define the standard deviation estimate in terms of 5 neighbouring spline sketches due to the slightly larger support of the original IRF for the face dataset, however one can easily construct equivalent closed form estimates for larger or smaller neighbourhoods\footnote{See Section \ref{Sec: Code} for details on MATLAB examples. }. As with the local mean estimates we need to use a background level estimate. Further details can be found in the code.

\section{Cram\'er-Rao Error Bounds}\label{SUBSEC: CRB}
In this section, the Cram\'er-Rao error bounds of the spline sketches for $p=0,1,2$ are compared with the original Fourier sketch as well as the full TCSPC histogram (i.e.\  no compression). The Cram\'er-Rao bound (CRB) gives a lower bound for the root mean squared error (RMSE) of an estimator $\hat{\theta}$ and can therefore provide the best case performance one could achieve through a reconstruction algorithm. Given the observation model $\pi$ in (\ref{Eqn: Observ model}) and the corresponding Fisher information matrix (FIM), defined as 
\begin{equation}
\label{eqn: Fisher info matrix}
    \mathcal{I}_{\text{data}}(\theta)\;=\; n \mathbb{E}_{x\sim\pi} \Bigg[\Bigg(\frac{\partial \log \pi(x \mid \theta)}{\partial\theta}\Bigg)^2\Bigg],
\end{equation}
then the optimal Cram\'er-Rao RMSE, in terms of the full data, is defined as 
\begin{equation}
    \label{eqn: Optimal MSE}
    \text{RMSE}_n\;=\;\sqrt{\sum^{2K}_{k=1} [\mathcal{I}_{\text{data}}(\theta)^{-1}]_{\{kk\}}}.
\end{equation}
Equivalently, for the sketched case the FIM is defined as (see for instance \cite{sheehan2021sketchedlidar})
\begin{equation}
\label{Eqn: Sketched FIM}
\big(\mathcal{I}_{\text{sketch}}(\theta)\big)_{ij}\; =\; n\dfrac{\partial\mathbf{z}_\theta}{\partial\theta_i}\Sigma^{-1}_{\theta}\dfrac{\partial\mathbf{z}_\theta}{\partial\theta_j},
\end{equation}
where for shorthand we denote $\mathbf{z}_\theta = \mathbb{E}_{x\sim\pi} \Phi_p(x)$ and where $\Sigma_{\theta}$ denotes the covariance matrix of the spline sketch. Similarly, we define the optimal sketched Cram\'er-Rao RMSE as 
\begin{equation}
    \label{eqn: Optimal Sketch MSE}
    \text{RMSE}_M\;=\;\sqrt{\sum^{2K}_{k=1} [\mathcal{I}_{\text{sketch}}(\theta)^{-1}]_{\{kk\}}}.
\end{equation}
\par In this section, the IRF is a Gaussian function with pulse width $\sigma$. This ensures that the formula in (\ref{eqn: Optimal Sketch MSE}) has a closed form solution, however, one can expect similar results to hold for more general IRFs. In addition, as many of the environmental factors need to be fixed, we choose ones that reflect typical lidar scenes. As the CRB is dependent on many factors, the following results vary one factor at a time so that one can understand the effect. 

\subsection{Spatially Averaged CRB Performance}
\begin{figure}[h!]
	\centering
	\includegraphics[width=0.48\textwidth]{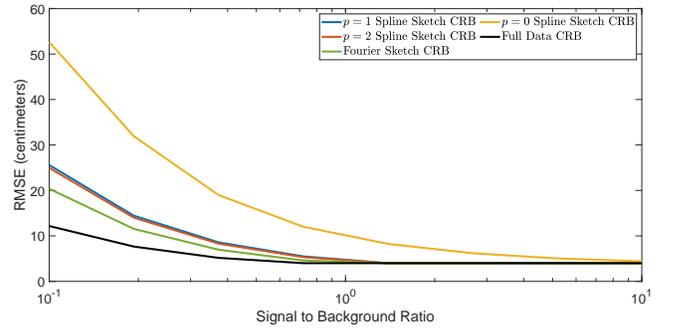}
	\caption{CRB for varying SBR for a Gaussian pulse width fixed at 64cm. Sketch size fixed at $M=8$.}
	\label{FIG: CRB wide pulse}
\end{figure}

\par We begin exploring the spatially averaged CRB performance of the different sketches. Each result is averaged along 1000 uniform depths across the timing depth window $[0,T-1]$. In our first experiment, we
vary the SBR from a low ($10^{-1}$) to a high ($10$) SBR regime. Here the acquisition time corresponds to a depth range $24$m where each of the $T$ timing bins equates to a precision of $4$cm. The sketch size for each of the spline and Fourier sketches is fixed at $M=8$ and the number of photon detections is $n=1000$. \Cref{FIG: CRB wide pulse} shows the CRB for the different sketches for a Gaussian pulse width of $\sigma = 64$cm for spline sketches of degree $p=0,1, 2$, a Fourier sketch and a full data TCSPC histogram (i.e.\  no compression). First, while the degree $p=0$ spline sketch (i.e. coarse binning) is able to achieve good performance at high SBR this reduces as the SBR gets less. Furthermore the $p=0$ RMSE is the highest of all the sketches. There is a significant reduction of RMSE as the degree of the spline is increased to $p=1$. Notably, the improvement of using a degree $p=2$ spline sketch is negligible in comparison to its linear counterpart. This is potentially due to the fact that locally first order statistics are almost sufficient in capturing the depth information and therefore the RMSE saturates for $p\geq 1$. There is a slight loss of information in comparison to using the original Fourier sketch of the same size, however one gains in the computational simplicity of constructing the spline sketches. Nonetheless, for a moderate SBR of 1, the RMSE of all the estimators apart from coarse binning are approximately the same at 5cm. 

\par
\begin{figure}[h!]
	\centering
	\includegraphics[width=0.48\textwidth]{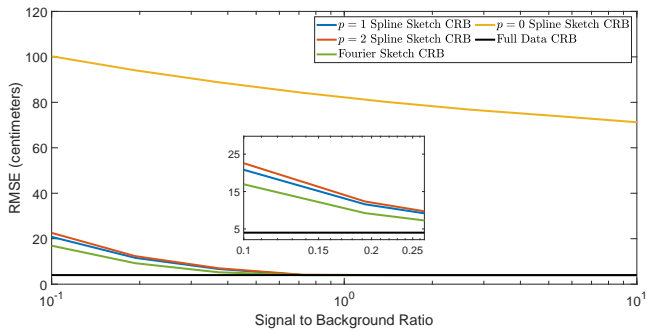}
	\caption{CRB for varying SBR for a Gaussian pulse width fixed at 24cm. Sketch size fixed at $M=8$.}
	\label{FIG: CRB narrow pulse}
\end{figure}

 \Cref{FIG: CRB narrow pulse} shows the results for the same set up as before but now with a narrower Gaussian pulse width of $\sigma=24$cm (as might occur in medium-high flux scenarios). In this case, one can clearly see that the coarse binning method fails catastrophically even at high SBR levels (the reason for this will become clear in the next subsection). In contrast, the other methods achieve similar performance to before and in fact are slightly more accurate. In both \Cref{FIG: CRB wide pulse} and \Cref{FIG: CRB narrow pulse}, we see that each of the $p=1,2$ degree splines and the Fourier sketches converge to the devices resolution (4cm in this case) as the SBR becomes high. 

\par
\begin{figure}[h!]
	\centering
	\includegraphics[width=.48\textwidth]{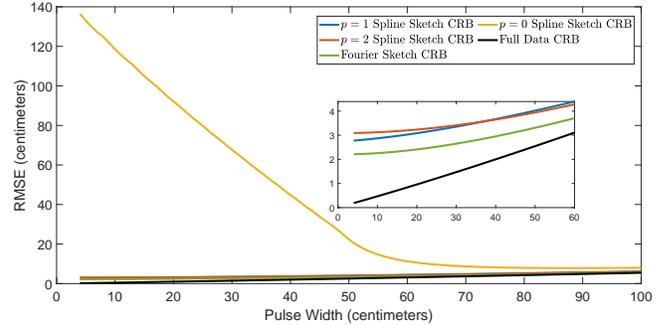}
	\caption{CRB for varying Gaussian pulse width with the SBR fixed at 1 and a sketch size fixed at $M=8$.}
	\label{FIG: CRB all pulse}
\end{figure}

 To further understand the relationship between the pulse width and the RMSE of the estimators, we fix the SBR at 1 and vary only the pulse width within the range of 2 and 100 cm. \Cref{FIG: CRB all pulse} demonstrates that the coarse binning RMSE is extremely dependent on the pulse width of the IRF, whilst, in contrast, the other methods are agnostic to the pulse shape. Some practitioners, for instance \cite{gyongy2020high}, have proposed increasing the pulse width of the device's laser to achieve a coarse binning RMSE that is only slightly worse than using either the sketches or full data. However, this is only practical in certain circumstances, and if the device experiences photon pile-up (see \Cref{SUBSEC: Range Walk Correction}) then the width of the detected signal can collapse towards a Dirac function. In such an event, coarse binning would fail catastrophically as illustrated in \Cref{FIG: CRB all pulse}.

\subsection{CRB Spatial Dependence}

\par
\begin{figure}[h!]
	\centering
	\includegraphics[width=.48\textwidth]{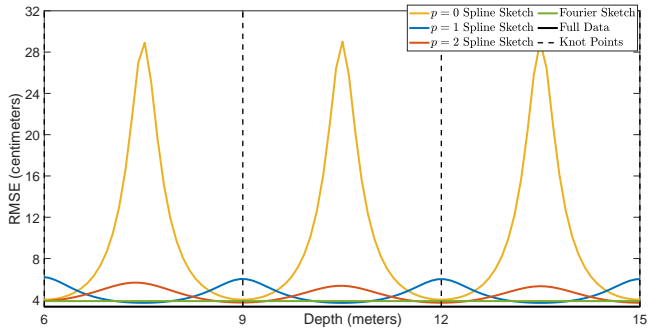}
	\caption{CRB for varying depth positions with the SBR fixed at 1, the Gaussian pulse width fixed at 64cm and a sketch size fixed at $M=8$.}
	\label{FIG: CRB temporal sweep}
\end{figure}

 By analysing the poor performance of coarse binning with narrow IRFs reveals a more troubling phenomenon. The results presented in the previous subsection averaged the CRBs over the whole timing depth window. However, the CRBs, with the exception of the Fourier sketch, do vary as a function of target depth. Therefore the average CRBs may be hiding a substantially worse performance at certain depths. We therefore next investigate the spatial dependency of the CRBs for the different sketches. To do so, the SBR and pulse width are fixed to 1 and 64cm, respectively, and the depth of the reflector is varied along the range of the lidar device. \Cref{FIG: CRB temporal sweep} shows the results. As expected, the CRBs for the full data and Fourier sketch are independent of the depth of the reflector. In contrast, the coarse binning estimator ($p=0$) varies from a small RMSE at the knot point to a significantly larger maximum error ($\sim 28$cm) at the centre of the knot interval. This can be intuitively understood as follows. If the return pulse straddles a knot point then its precise position can be estimated accurately from the relative magnitudes of the neighboring histogram bins. However, if the pulse lies solely in one knot interval then the estimator is blind to where the pulse is within the knot interval. This dependency also occurs to a certain extent for $p=1,2$ degree spline sketches but at a much dampened state. Interestingly, the $p=1$ spline sketch achieves the largest RMSE at the knot points and the smallest RMSE at the centre of the knot intervals which is in contrast to the even degree splines sketches of $p=0,2$. This result importantly highlights the worst case performance one can expect to achieve from each spline sketch which, as demonstrated, can be extreme for coarse binning yet is controllable for both the linear and quadratic spline sketches.

\section{Simulations}\label{SEC: Experiments}
In the section we evaluate the proposed algorithms on both synthetic and real datasets and compare their performance to the CRB discussed in \Cref{SUBSEC: CRB}.

\subsection{Algorithm Dependent Performance}\label{APP: Compare to CRB}
\begin{figure}[h!]
	\centering
	\includegraphics[width=.48\textwidth]{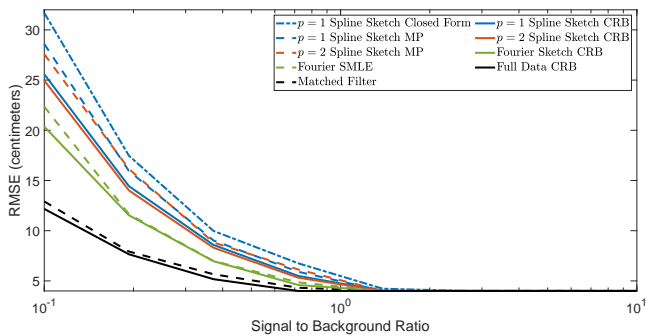}
	\caption{A comparison of the proposed reconstruction algorithms and their respective CRBs with the SBR fixed at 1, the Gaussian pulse width fixed at 64cm and a sketch size fixed at $M=8$.}
	\label{FIG: Actual_vs_CRB}
\end{figure}

While the CRBs provide us with performance bounds for any estimator using the associated statistics, it is also necessary to understand how a practical algorithm performs with respect to the associated CRB. We therefore conclude this section by showing how the different algorithms presented in \cref{SEC: Spline Sketch Reconstruction} compare against the CRBs. The set up for this experiment is identical to the one presented in \Cref{FIG: CRB wide pulse}. We vary the SBR from a $10^{-1}$ to $10$. The depth range is $24$m with $T$ timing bins equating to a precision of $4$cm each. The sketch size is $M=8$, the number of photon detections is $n=1000$, and the IRF is modelled as a Gaussian pulse with width $\sigma = 64$cm. \Cref{FIG: Actual_vs_CRB} shows the that all algorithms perform equivalently at high SBR.
While for low SBR all algorithms exhibit a small drop in performance compared with their associated CRB. The MP algorithm appears to perform similarly for both $p=1$ and $p=2$ while the closed form estimate is slightly worse. This is most probably due to the assumption that the support of the IRF does not exceed a single know interval. This could be remedied by additional fine-tuning as discussed in (\ref{Eqn: closed form depth estimates}).

\subsection{Real Dataset}\label{SUBSEC: Real Data}
\begin{figure*}[t!]
	\centering
	\includegraphics[width=1\textwidth]{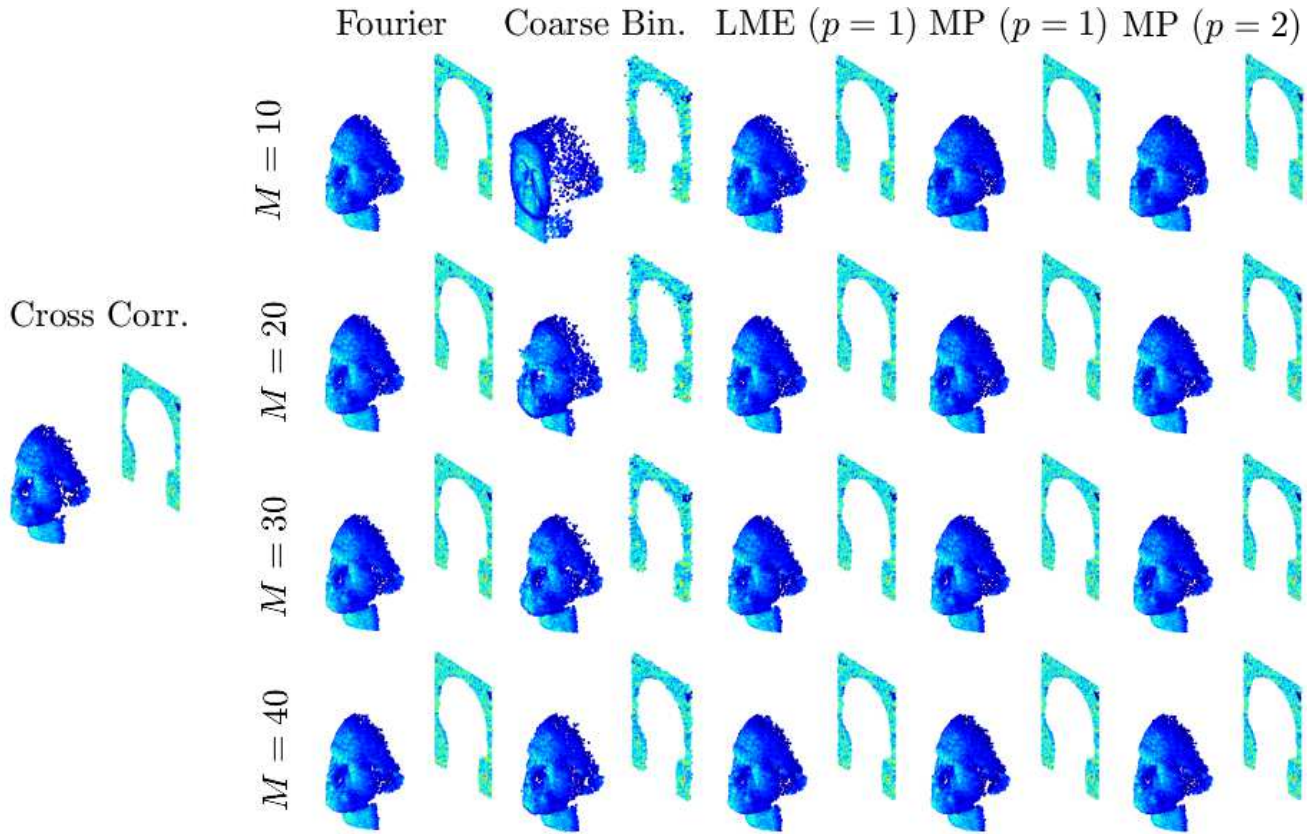}
	\caption{Reconstruction of the face dataset using a Fourier sketch, coarse binning ($p=0$), the local means estimator in \Cref{SUBSEC: Closed Form Sol}, and the $p=1,2$ spline sketches using the MP algorithm.}
	\label{FIG: Face Recon}
\end{figure*}

We evaluate the proposed algorithms in \Cref{SEC: Spline Sketch Reconstruction} on a real dataset consisting of a polystyrene head placed at a distance of 40 metres\cite{altmann2016lidar}. We compare the proposed methods with the cross correlation algorithm\cite{McCarthy:09} that works directly on the full TCSPC (i.e.\  no compression), as well as the SMLE algorithm in \cite{sheehan2021sketchedlidar} that uses Fourier features to construct the sketch. The polystyrene head dataset has size of $141\times 141$ pixels with $T=4613$. Most of the pixels in this scene contain exactly $K=1$ surface and the average number of photon detections per pixel is 337. \Cref{FIG: Face Recon} and \cref{TABLE: Face RMSE scores} show the reconstruction of the scene and the corresponding RMSE metrics, respectively, using a Fourier sketch and the SMLE algorithm in \cite{sheehan2021sketchedlidar}, coarse binning ($p=0$), the closed form solution in \Cref{SUBSEC: Closed Form Sol}, and the $p=1,2$ spline sketches using the MP algorithm in \Cref{SEC: Spline Sketch Reconstruction}. These compression algorithms are compared to the matched filtering algorithm \cite{McCarthy:09} that acts on the original uncompressed TCSPC histogram. For the compression techniques, we consider a sketch size of $M=10,20,30$ and 40. Initially, we see that the coarse binning method fails to accurately recover the scene for the smaller sketch sizes. In comparison, both $p=1,2$ degree splines manage to retain enough information from the ToF data to allow for accurate reconstruction. One can observe a slight drop in performance between the closed form solution and the matching pursuit algorithm especially around the border of the head. Interestingly, a linear and quadratic spline sketch of just $M=20$ allows for very accurate reconstruction in comparison to both the Fourier sketch and the full data reconstruction. For this specific dataset, the linear and quadratic spline sketches correspond to a compression of approximately $95\%$ in comparison to storing the raw photon detection times without sacrificing the overall quality of the reconstructed image.

\begin{table*}[htp]
\label{TABLE: Face RMSE scores}
\begin{tabular}{|c|c|c|c|c|c|c|}
\hline
Sketch Size $M$ & \multicolumn{1}{l|}{Cross Corr.} & Fourier & Coarse Bin. & LME ($p=1$) & Spline MP ($p=1$) & Spline MP ($p=2$) \\ \hline
10               &             & 8.2     & 74.5        & 15.3                          & 12.1                     & 11.7                      \\ \cline{1-1} \cline{3-7} 
20              &     4.4                              & 6.2     & 22.8        & 11.4                          & 8.4                      & 8.5                      \\ \cline{1-1} \cline{3-7} 
30              &                                  & 4.8     & 18.1        & 8.6                          & 6.2                      & 6.4                      \\ \cline{1-1} \cline{3-7} 
40              &                                  & 4.6     & 15.1        & 7.0                          & 5.7                      & 5.9                      \\ \hline
\end{tabular}
\caption{RMSE (bins) of the different compression techniques for a sketch of size $M=6,12$ and 24, respectively.}
\end{table*}

\subsection{Range Walk Correction Simulations}
\begin{figure}[h!]
	\centering
	\includegraphics[width=.49\textwidth]{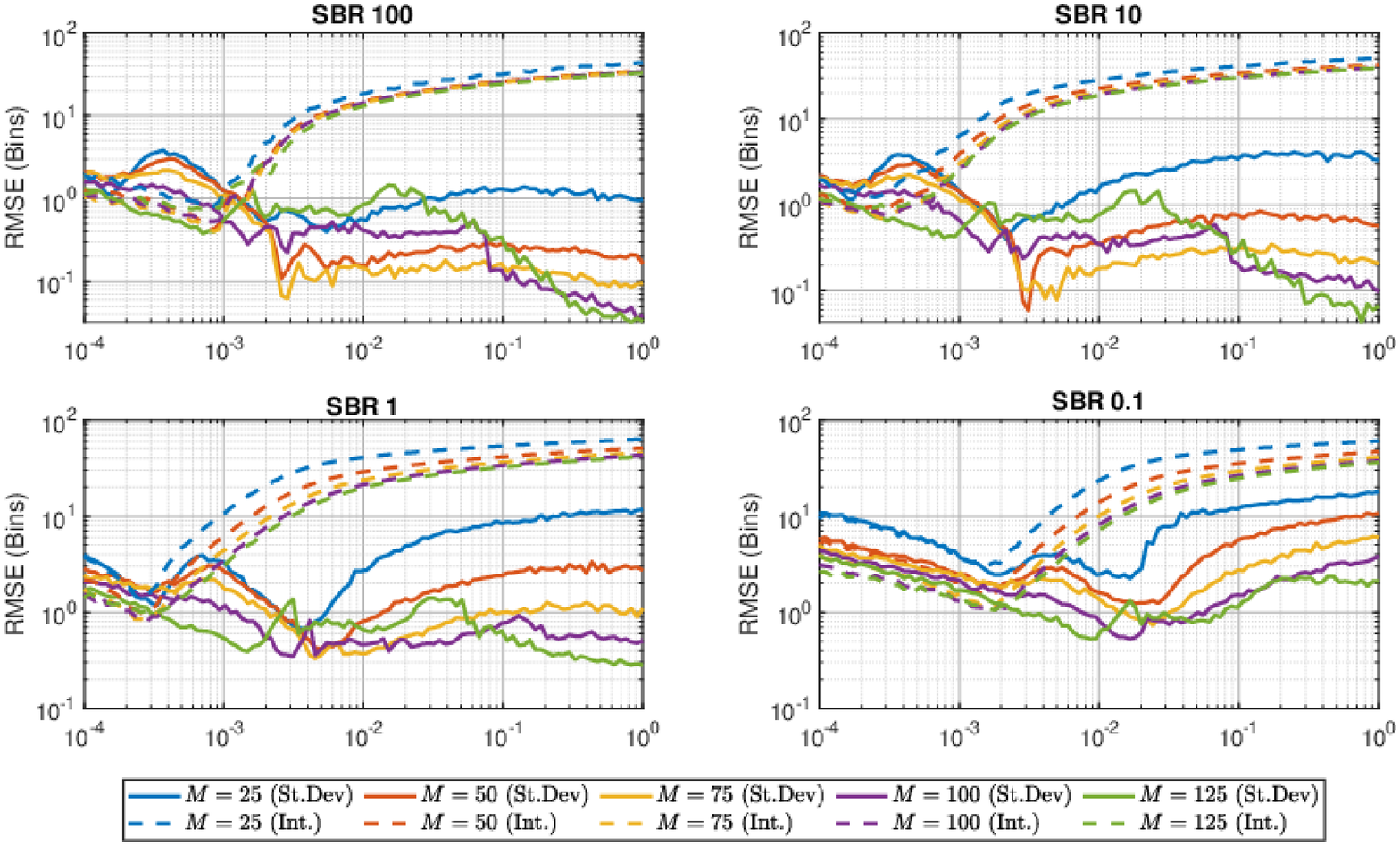}
	\caption{The RMSE of both the intensity and shape based range walk correction models for different SBR levels.}
	\label{FIG: RW RMSE}
\end{figure}

To evaluate our shaped based range walk model discussed in \Cref{SUBSEC: Range Walk Correction}, we use the IRF in\cite{altmann2016lidar} where $T=4613$ (see \Cref{SUBSEC: Real Data} for more details) and build a look-up table as in \Cref{FIG: RW LUT tables for both} for a sketch size of $M=25,50,\dots, 125$ using the model in \cite{heide2018sub}. Once an intensity or standard deviation value has been estimated from the spline sketches, we consult the aforementioned look-up tables using the estimates to correct for the range walk. For 50 different target reflectivities (on a log-scale between $10^{-4}$ and 1) and SBR values of 100,10,1 and 0.1, we compute the root mean squared error (RMSE) calculated over 250 random target distance initialisations. \Cref{FIG: RW RMSE} shows the RMSE for each of the SBR values as a function of target reflectivity. One can initially see that as the point of intensity saturation is hit, the RMSE grows rapidly as the intensity based range walk model is unable to accurately distinguish the correct range walk. In contrast, the shape based range walk model stays at quite a constant error level throughout despite the increase in target reflectivity. For the lowest SBR of 0.1, the smaller sketch sizes of $M=25$ and $M=50$ have a larger RMSE for the higher target reflectivity. This is because in challenging imaging scenes, a larger sketch size is required to accurately estimate and remove the contribution from background sources. These results demonstrate additional statistical information captured by the quadratic spline sketch provide further benefits in complex environments.

\section{Conclusion}\label{SEC: Conclusion}
In this paper, we proposed a semi-local spline sketch that can be efficiently implemented in hardware and that achieves the compression capabilities of the original Fourier sketch whilst enjoying minimal overhead computations on-chip. Simulations demonstrated that both the linear and quadratic splines sketches achieve almost the same performance as the original sketch with only a slight drop in performance for very low SBR scenes. Increasing the degree of the spline sketch to $p=2$ also enabled us to capture local 2nd order statistics that allow us to accurately correct for range walk error where intensity based range walk models fail to do so. 
\section{Code Availability}\label{Sec: Code}
 A MATLAB implementation of all the algorithms discussed are available at the repository \texttt{\url{https://gitlab.com/mpsheehan1995/spline-sketch-lidar}}.

\bibliographystyle{IEEEtran}
\bibliography{bibliography}

\end{document}